\documentclass[twocolumn,showpacs,preprintnumbers,amsmath,amssymb]{revtex4}

\usepackage{color}
\usepackage{graphicx}
\usepackage{dcolumn}
\usepackage{natbib}
\usepackage{bm}
\def\be{\begin{equation}}
\def\ee{\end{equation}}
\def\bea{\begin{eqnarray}}
\def\eea{\end{eqnarray}}

\def\nn{\nonumber}

\begin{document}

\title{\bf Measurement-Induced Phase Transition in a Quantum Spin System}

\author{Shrabanti Dhar and Subinay Dasgupta} 
\affiliation{
Department of Physics, University of Calcutta,
92 Acharya Prafulla Chandra Road, Kolkata-700009, India}

\begin{abstract}
 Suppose a quantum system starts to evolve under a Hamiltonian from some initial state. When for the first time, will an observable attain a preassigned value? To answer this question, one method often adopted is to make instantaneous measurements periodically and note down the serial number for which the desired result is obtained for the first time. We apply this protocol to an interacting spin system at zero temperature and show analytically that the response of this system shows a non-analyticity as a function of the parameter of the Hamiltonian and the time interval of measurement. In contrast to quantum phase transitions, this new type of phase transition is not a property of the ground state and arises from the Hamiltonian dynamics and quantum mechanical nature of the measurement. The specific system studied is transverse Ising chain and the measurement performed is, whether the total transverse magnetic moment (per site) is not equal to 1. The results for some other types of measurement is also discussed. 
 \end{abstract}
\pacs{03.65.Ca, 03.65.Ta, 64.70Tg }    
\maketitle

{\bf Introduction:} Classical first passage problem is easy to formulate. We allow a system to undergo time evolution starting from some initial state, and check when a dynamical variable attains some preassigned value $X$  for the first time. The corresponding quantum mechanical version is difficult to handle \cite{Muga}
  because measurement of time is associated with fundamental issues and because continuous projective measurement prevents time evolution of a quantum system (the quantum Zeno effect \cite{QZE}). Following a prescription by Misra and Sudarshan \cite{Misra} one may perform \cite{own1} at interval of time $\tau$, instantaneous projective measurements which answers the  the question {\em `Is the value of the observable in question equal to $X$?'}. We then identify the probability that the answer is {\emph {yes}} for the first time at the $n$-th measurement, as the probability that our measurement will give $X$ for the first time at time $n\tau$ ($n= 1, 2, \cdots$). This gives an experimentally feasible protocol for measuring the probability of first occurrence within the framework of the measurement hypothesis of quantum mechanics. Such periodic measurements have also been proved to be relevant for quantum communication\cite{BO}\\
  
Since a quantum measurement alters a system due to its projective nature, it is interesting to investigate the features of the changes brought into a system by measurement itself. It has been demonstrated experimentally that projective measurement can generate entanglement between a pair of solid-state qubits \cite{pfaff} and between a pair of superconducting qubits \cite{Roch}.  Theoretical investigations \cite{mazzucchi,SCQubit,qst,banerjee} on various systems has also been performed.    
 In this article, we show that when such a study is performed on an interacting quantum Ising system, the response of the system shows non-analytic behaviour as a function of the parameter of the Hamiltonian and the time interval of measurement. 
This is a new type of phase transition, which does not arise from any singular behaviour of the ground state, but is a result of the quantum mechanical nature of the measurement and dynamics.

Specifically, we consider a periodic transverse Ising chain of $N$ Ising spins ($s=\pm 1$) at zero temperature described by the Hamiltonian
\be \mathcal{H} =   -  \sum_{j=1}^N s^x_j s^x_{j+1} - \Gamma \sum_{j=1}^N s^z_j \label{H_def} \ee
 and perform on it a measurement (at interval of time $\tau$) that provides binary answer to the question 
 \be \mbox{\em {Is the magnetic moment (per site) $M_z \ne 1$?}} \label{Question1} \ee
 Here $\Gamma$ is the strength of the transverse field and $M_z =(1/N) \sum_{i=1}^N s_i^z  $ is the transverse magnetic moment per site. One can calculate analytically the probability of first occurrence $p_n$ that the result will be {\emph yes} for the first time at the $n$-th measurement, for an arbitrary value of $\tau$ (not necessarily small) and find that this quantity decays exponentially with $n$:
 \be p_n  \sim  e^{-\beta n} \label{pn_def_beta} \ee
 where $\beta$ is independent of $n$ but is a function of $\tau$ and $\Gamma$. We shall see below that $\beta \propto N$ for large $N$ and that in the limiting case of small $\tau$, $\beta \propto \tau^2$. Hence it is convenient to introduce a quantity
 \[ \alpha = \frac{\beta}{N\tau^2} \]
 which we shall call the {\em decay constant}. In terms of $\alpha$,  the variation of $p_n$ will be given by
  \be p_n  \sim  e^{-N\tau^2 \alpha n}  \label{pn_def} \ee
 The central result of this work is that this decay constant shows non-analytic behaviour as a function of the field strength $\Gamma$ and time interval $\tau$. Thus, for a given $\tau > \pi/4$, the decay constant $\alpha$ increases with $\Gamma$ till a critical value $\Gamma_0$ and for $\Gamma > \Gamma_0$, it decreases with $\Gamma$. A similar 
 behaviour is also observed in the variation of $\alpha$ as a function of $\tau$ for a given value of $\Gamma < 1$. The discontinuity in the slopes $(\partial \alpha /\partial \Gamma)_{\Gamma_0}$ and $(\partial \alpha /\partial \tau)_{\tau_0}$ can be calculated in closed form. In view of the presence of non-analyticity, we call this change in behaviour a `phase transition', which indeed has no connection with the order-disorder transition at $\Gamma=1$ of the ground state of the Hamiltonian $\mathcal{H}$. We shall also discuss the cases of other types of measurement and report that when the measurement (\ref{Question1}) is replaced by `{\emph {Is $M_z=0$?}}', according to numerical studies with $N\le 24$ although $p_n$ still decays more or less exponentially with $n$,  there is no sign of any non-analytic behaviour of the decay constant.
 
{\bf Related works and background:}  There exists a vast literature for first passage problem and the measurement of time of arrival in quantum mechanics (for review, see \cite{Muga} and references in \cite{own1}). The  measurement protocol followed here was developed in two earlier papers \cite{own1} for measuring experimentally the time of arrival of a free quantum particle moving on a lattice under a tight-binding Hamiltonian. This analytic treatment was applicable for {\em small but finite} values of $\tau$. 

 As mentioned above, projective measurement on a spin system has been studied recently in various contexts \cite{pfaff,Roch,mazzucchi,SCQubit,qst,banerjee}. However, to our knowledge, long range order has not yet been investigated.

{\bf Analytic expression for the probability of first occurrence:} One can solve \cite{LSM_etc} the transverse Ising Hamiltonian in Eq. (\ref{H_def}) for arbitrary $N$ by transforming it to a 
 direct sum of commuting Hamiltonians $\mathcal{H}_k$ given by
 \begin{align}
 \mathcal{H}_k = (-2 i \sin k) \left[ c_k^{\ast}c_{-k}^{\ast} + c_k c_{-k} \right]  ~ \nonumber \\
 \;\;\;\;\;\;\;\;\;\;\;\;\;\;  - 2(\Gamma + \cos k)\left[ c_k^{\ast}c_k + c_{-k}^{\ast}c_{-k} - 1 \right] \label{Hk_def} 
\end{align} 
  where $k=(2\ell+1)\pi/N$, with $\ell=0,1,\cdots,\frac{N}{2}-1$ and $c_k^{\ast}$, $c_k $ are Fermion creation and destruction operators.  
  In the even-occupation space, the two eigenstates of $\mathcal{H}_k$ are the ground state $|{\rm GS}_k\rangle$ and the excited state $|{\rm ES}_k\rangle$ with respective eigenvalues $-\lambda_k$ and $\lambda_k$ given by,
\bea   \lambda_k &  =  & 2\sqrt{\Gamma^2 +1 +2\Gamma \cos k} \label{Lambda} \\
|{\rm GS}_k\rangle & = & i \cos \theta_k |11_k\rangle - \sin \theta_k |00_k\rangle \label{GS} \\
|{\rm ES}_k\rangle & = & i \sin \theta_k |11_k\rangle + \cos \theta_k |00_k\rangle \label{ES} \\
\tan \theta_k & = & \dfrac{-\sin k}{\Gamma + \cos k + \sqrt{\Gamma^2 +1 +2\Gamma \cos k}} \label{theta} \eea
Here $|11_k\rangle$ ($|00_k\rangle$) is a state that has a (no) fermion at $+k$ and $-k$. 
Let us denote the possible (even-occupation) states in $k$-space as $|b\rangle = |\cdots (-k, k) \cdots \rangle $ where  $(-k, k)$ corresponds to $|11_k\rangle$ or $|00_k\rangle$. There will be $2^{N/2} \equiv N'$ such states, which we denote by $|b_1\rangle, |b_2\rangle,  \cdots |b_{N'}\rangle$ with $|b_1\rangle$ as the state with no fermion at any $k$ and $|b_{N'}\rangle$ is the state with a fermion at every $\pm k$. 
Note that, since $M_z=(1/N)\sum_i s_i^z = (1/N) \sum_{k=-\pi}^{\pi} (2c_k^{\ast} c_k - 1)$, we must have $\langle b_{N'} |M_z |b_{N'}\rangle = 1$.

\begin{figure*}
{\includegraphics[clip,width=8.8cm, angle=0]{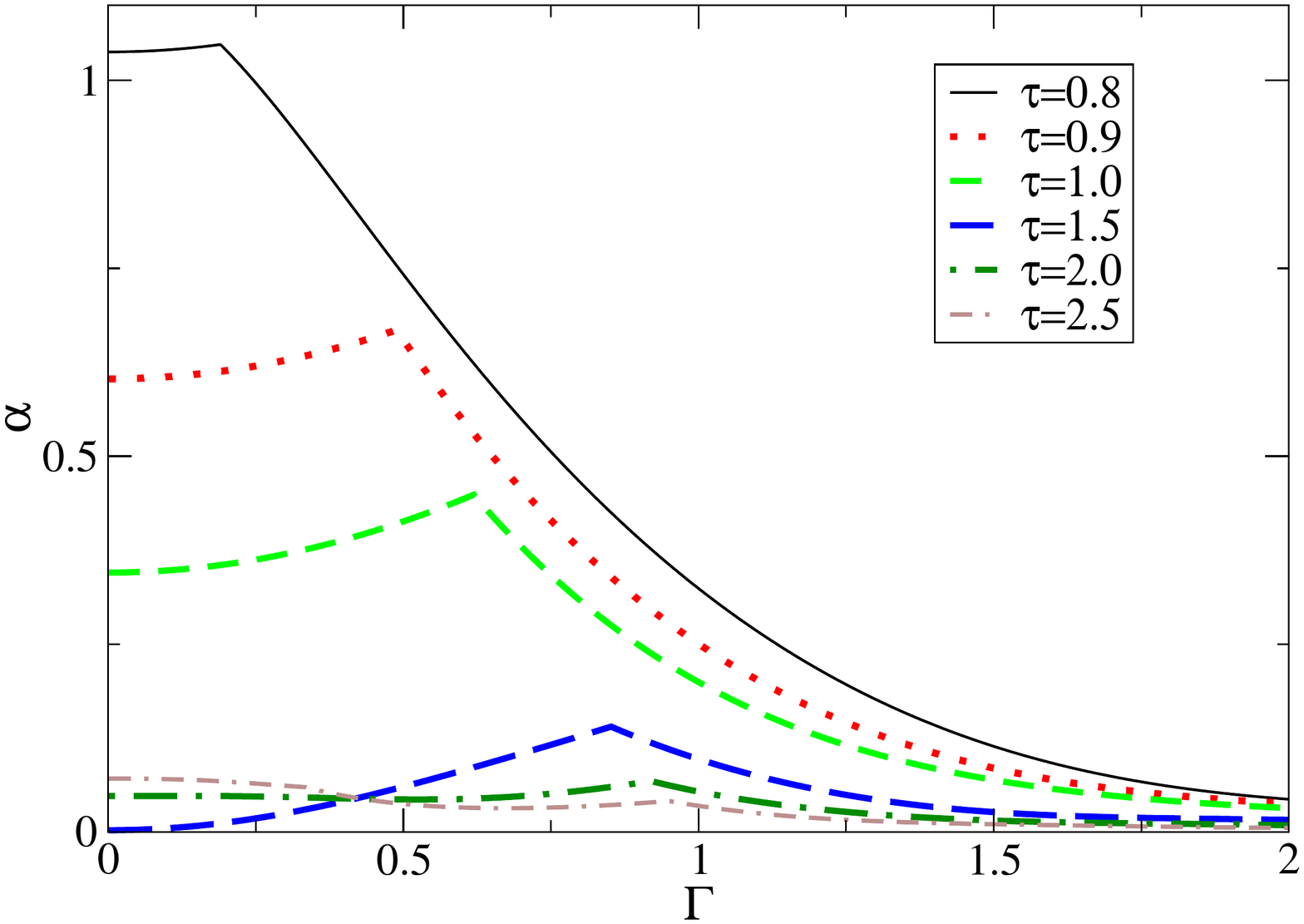}}
{\includegraphics[clip,width=8.8cm, angle=0]{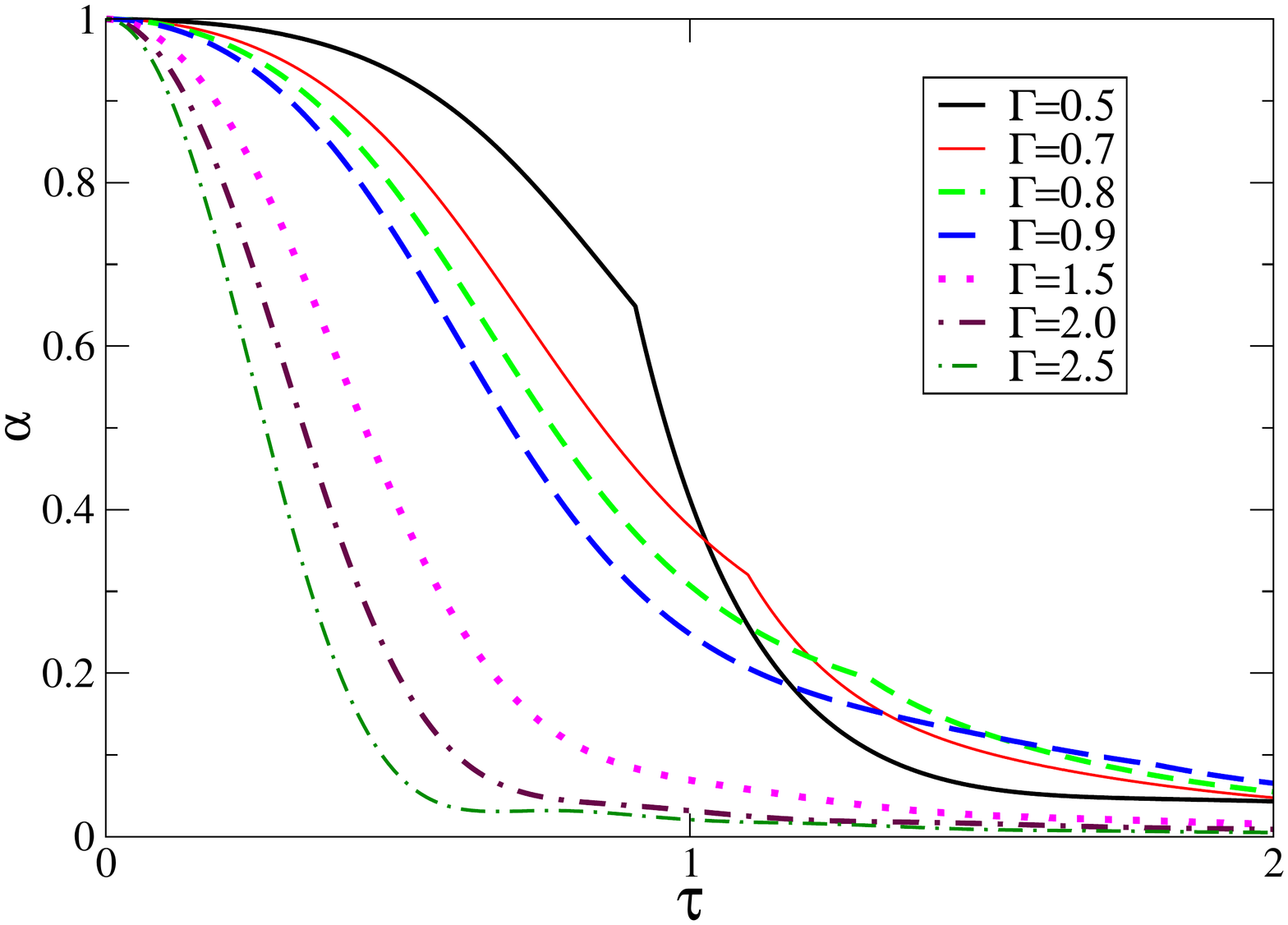}}
\caption{(color online)  Decay constant $\alpha$, as given by Eq. (\ref{alpha}), with respect to $\Gamma$ at different $\tau$ (left panel) and with respect to $\tau$ at different $\Gamma$ (right panel). Note that for some of the $\tau$ values in the left panel , the slope $(\partial \alpha /\partial \Gamma)$ changes discontinuously at some critical value $\Gamma_0$ and similarly in the right panel, for some of the $\Gamma$ values $(\partial \alpha /\partial \tau)$ changes discontinuously at some critical value $\tau_0$. }
\label{alpha_fig}
\end{figure*}

The algorithm we follow is this : We start from an initial configuration of the system $|\psi (0)\rangle$ at $t=0$, let it evolve under the Hamiltonian  $\mathcal{H}$ of Eq. (\ref{H_def}) for time $\tau$; then we perform the measurement of (\ref{Question1}); then again let the system evolve for time $\tau$ and make the same measurement, and so on.
The first occurrence probability at the $n$-th measurement is given by,
\be p_n = \sum_{j\ne N'} |\langle b_j |\psi (n\tau - \epsilon)\rangle|^2 \ee
where $|\psi (n\tau - \epsilon)\rangle$ is the wave function just before the $n$-th measurement and $\epsilon$ is a small positive number.
The wave function after the $n$-th measurement will be
\begin{align}
 |\psi(n\tau + \epsilon) \rangle &=|\psi(n\tau - \epsilon) \rangle - \sum_{j\ne N'} \langle b_j | \psi(n\tau - \epsilon) \rangle |b_j \rangle \nn \\
 &=\langle b_{N'} | \psi(n\tau - \epsilon) \rangle |b_{N'} \rangle \label{collapse} 
\end{align}
Thus all the states with $M_z \ne 1$ are projected out and only the state $|b_{N'} \rangle$ (which is incidentally not an eigenstate of the Hamiltonian) is retained. Let the amplitude of the wave function after the $n$-th measurement be $C_n$, so that
\be |\psi(n\tau + \epsilon)\rangle = C_n |b_{N'} \rangle \label{amplitude} \ee
and the probability that the result is {\em no} for all the first $n$ measurements is $|C_n|^2$. Using Eqs.(\ref{GS},\ref{ES}) we can express $|11_k\rangle$ in terms of $|{\rm GS_k}\rangle$ and $|{\rm ES_k}\rangle$ and obtain the wave function after time evolution as,
\be |\psi([n+1]\tau - \epsilon)\rangle =   C_n \bigotimes  \left( A_k |11_k \rangle + B_k |00_k \rangle \right) \label{def_AB} \ee
with $\hbar=1$, $A_k = \left(e^{-i\lambda_k \tau} \sin^2 \theta_k  +  e^{i\lambda_k \tau} \cos^2 \theta_k \right) $ and $B_k = -\sin(\lambda_k \tau) \sin(2\theta_k)$.
Hence  
\be C_{n+1}/C_n = \prod_k  A_k \label{ratioC} \ee 
This equation is quite general in the sense that it is true for all values of $n$ and $\Gamma$. Noting that $C_{n+1}/C_n$ is independent of $n$, and that $p_{n+1} = |C_n|^2 - |C_{n+1}|^2 $, 
the ratio of first occurrence probabilities at two successive measurements is,
\be \frac{p_{n+1}}{p_n} = \prod_k  |A_k |^2  \label{ratio1}\ee
Finally,
\[ p_n = p_0 \exp\left(n\sum_k\log\left[1-g_k^2 \right]\right)\]
with $g_k  =   [2 \sin k \,  \sin (\lambda_k \tau)]/\lambda_k$ using the equality $\sin (2\theta_k) = - (2 \sin k) /\lambda_k$. Here $p_0$ is the proportionality constant independent of n. According to Eq. (\ref{pn_def}), the decay constant is then given by,
\be  \alpha (\Gamma, \tau) = -\frac{1}{2\pi \tau^2} \int_0^{\pi} dk \, \log (1 - g_k^2)   \label{alpha} \ee
in the limit $N\rightarrow\infty$. Thus, $\alpha$ is independent of $N$ and the exponent $\beta$, as introduced in Eq. (\ref{pn_def_beta}) is proportional to $N$. Also, 
for small $\tau$, $\alpha  \approx 1$, and hence $\beta \propto \tau^2$, as mentioned above. The crucial result of our work is to observe that for larger $\tau$, the decay constant $\alpha$ shows interesting behaviour (Fig.~\ref{alpha_fig}) as the system now gets enough time between two successive measurements to generate new states.

The decay constant $\alpha$ shows non-analytic behaviour as a function of $\Gamma$ and $\tau$ at some `critical' points $(\Gamma_0, \tau_0)$ due to the logarithmic singularity in the expression (\ref{alpha}) for $\alpha$.  The critical point occurs when the argument of the logarithm vanishes :
\be \tau_0 \sqrt{1-\Gamma_0^2} = \frac{\pi}{4}, \;\;\;\;\;\; k_0=\cos^{-1}(-\Gamma_0) \label{critical}  \ee
Note that this equation is true only for $\Gamma_0 < 1$.
Let us first consider the case when variation of $\alpha$ as a function of $\Gamma$ is studied for a fixed $\tau$. Close to the critical point the quantity $(1-g_k)$ approaches zero as,
\begin{align}
\hspace*{-4.0 cm}1 - g_k = \frac{1}{2} \left( 1 + \frac{\pi^2}{4}\cot^2 k_0 \right) (k - k_0)^2 - \nonumber \\
\frac{1}{\sin k_0} (k - k_0)(\Gamma - \Gamma_0)
+ \frac{1}{2\sin^2 k_0} (\Gamma - \Gamma_0)^2 \label{g_Taylor1} 
\end{align}
Noting that for the integral in Eq. (\ref{alpha}) only the region around $k=k_0$ is important, and using the standard equality \cite{formula},
\[\lim_{a \to 0} \frac{a}{a^2 + x^2} = \pi \delta(x) {\rm sign} (a) \]
(where ${\rm sign} (a) = \pm 1$ according as $a$ is positive or negative), one gets 
the change in slope at the critical point as,
\be \Delta_{\Gamma} \equiv \left(\frac{\partial \alpha}{\partial \Gamma} \right)_{\Gamma_0^{+}}  - 
\left(\frac{\partial \alpha}{\partial \Gamma} \right)_{\Gamma_0^{-}}  =  -\frac{16 \sqrt{16\tau^2 - \pi^2}}{\pi \tau ( 4 - \pi^2 + 16\tau^2)} \ee

In a similar way, if we consider the variation of $\alpha$ as a function of $\tau$ for a fixed $\Gamma$ close to the critical point, the quantity $(1-g_k)$ can be expanded as, 
\begin{align}
\hspace*{-6.0cm}1 - g_k = \frac{1}{2} \left( 1 + \frac{\pi^2}{4}\cot^2 k_0 \right) (k - k_0)^2 + \nonumber \\ \pi \cos k_0 (k - k_0)(\tau - \tau_0) 
 + 2 \sin^2  k_0 (\tau - \tau_0)^2 \label{g_Taylor2} 
\end{align}
and the change in slope at the critical point becomes
\be \Delta_{\tau}  \equiv \left(\frac{\partial \alpha}{\partial \tau} \right)_{\tau_0^{+}}  - 
\left(\frac{\partial \alpha}{\partial \tau} \right)_{\tau_0^{-}} 
= -\frac{256 (1 - \Gamma^2)^{5/2}}{( 4 + (\pi^2 - 4) \Gamma^2)\pi^2 } \ee

{\bf Two other measurements - analytic treatment:} It is interesting to consider two other types of measurement for which one can calculate the decay constant analytically.  (i) Let us assume that the  effect of measurement is to retain some single state $|b_j \rangle$ (without worrying about the physical possibility of such measurement). Since $ |11_k \rangle$ after evolution over time $\tau$ becomes $( A_k |11_k \rangle + B_k |00_k \rangle)$ (as mentioned in Eq. (\ref{def_AB})) and $ |00_k \rangle$ after such evolution becomes $( A_k^{\ast} |00_k \rangle - B_k |11_k \rangle)$, the ratio of probabilities is again given by Eq. (\ref{ratio1}) and the decay constant $\alpha$ remains the same. (ii) If we replace the measurement (\ref{Question1}) by `{\em Is $M_z \ne \pm 1$ ?}'
 then the state after $n$-th measurement can be written as,
 \[ |\psi(n\tau + \epsilon)\rangle = C_n |b_1 \rangle + C'_n |b_{N'} \rangle  \]
 where $C_n$ and $C'_n$ are the complex amplitudes. (Note $\langle b_1 |M_z |b_1\rangle = -1$.)
 Using Eqs.(\ref{GS},\ref{ES}), one can again express $|00_k\rangle$ and $|11_k\rangle$ in terms of $|{\rm GS_k}\rangle$ and  $|{\rm ES_k}\rangle$ and obtain the wave function after evolution over time $\tau$, and therefrom the wave function after $n$-th measurement. Finally, the ratio of first occurrence probabilities is,
 \be \frac{p_n}{p_{n-1}} = \prod_k  \left(1 - g_k^2  \right)   +  \prod_k  \left(g_k^2  \right)   \label{ratio2} \ee
 giving a decay constant
\be \alpha_1  =  \alpha -\frac{1}{N \tau^2} \log \left( 1 +  \prod_k  \frac{g_k^2}{1 - g_k^2} \right) \label{alpha_1} \ee
We have checked numerically that the product on the right hand side is much less than unity in the region of our interest.  Hence, the behaviour of the decay constant  remains practically the same as that for measurement (\ref{Question1}).
\begin{figure}[!h]
{\includegraphics[clip,width=8.8cm, angle=0]{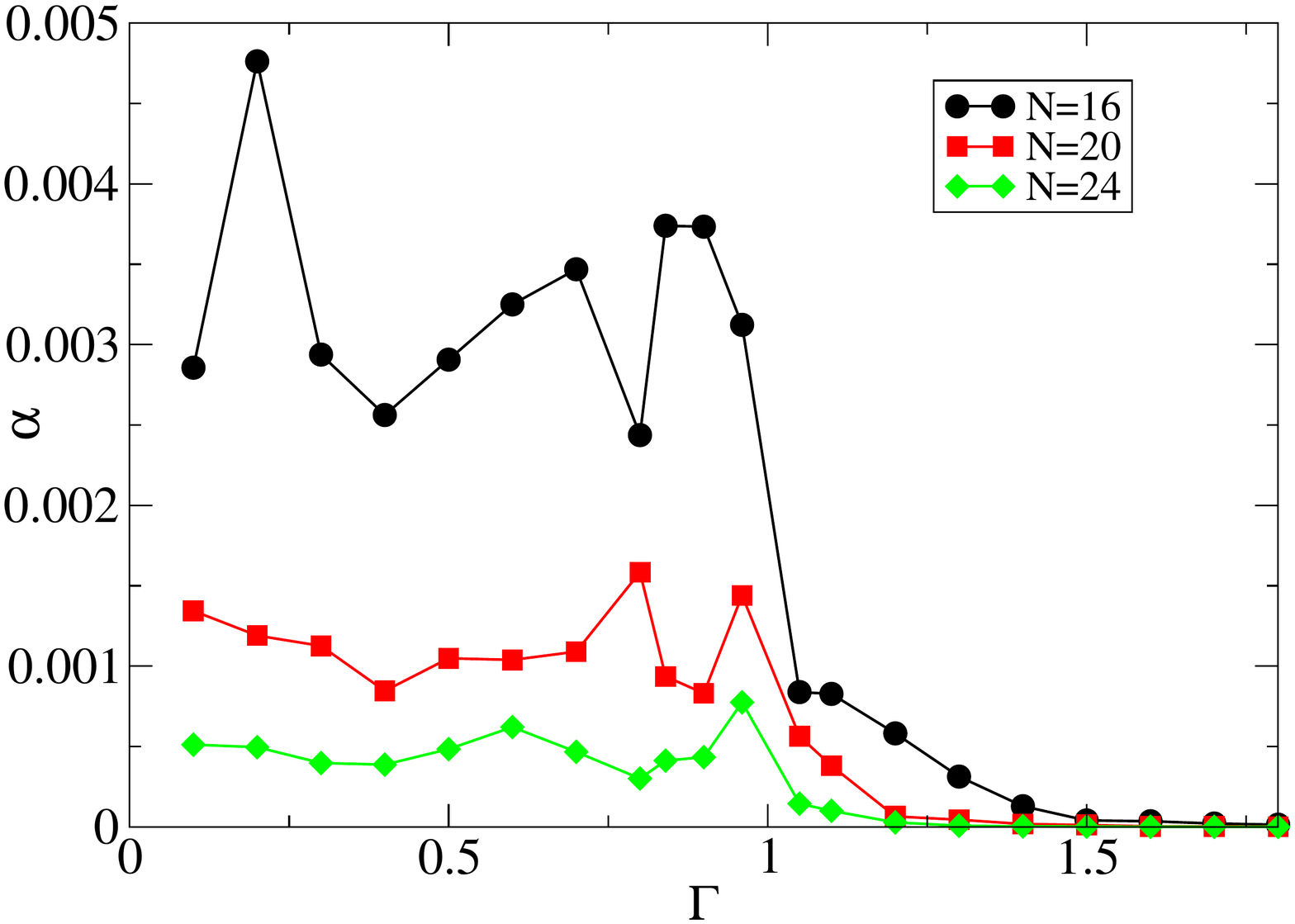}}
\caption{(color online) Decay constant $\alpha$ as a function of applied transverse field $\Gamma$ for the measurement (\ref{Question2}). Here $N=20$, $\tau=1.0$ and the initial state corresponds to the ground state. We have checked that the results are grossly independent of the choice of initial state. }
\label{alpha_IQ=0}
\end{figure}

\begin{figure*}
{\includegraphics[clip,width=8.8cm, angle=0]{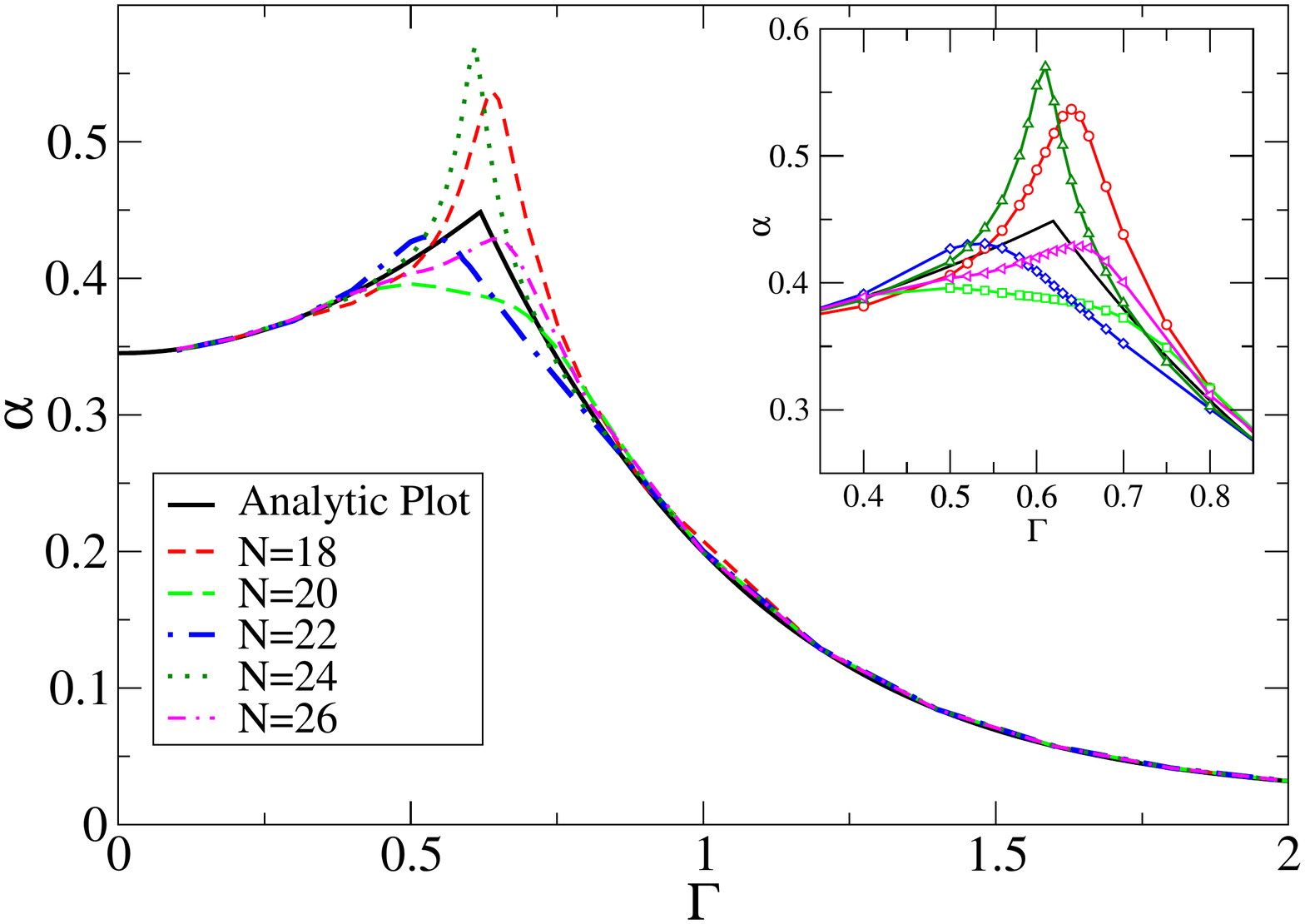}}
{\includegraphics[clip,width=8.8cm, angle=0]{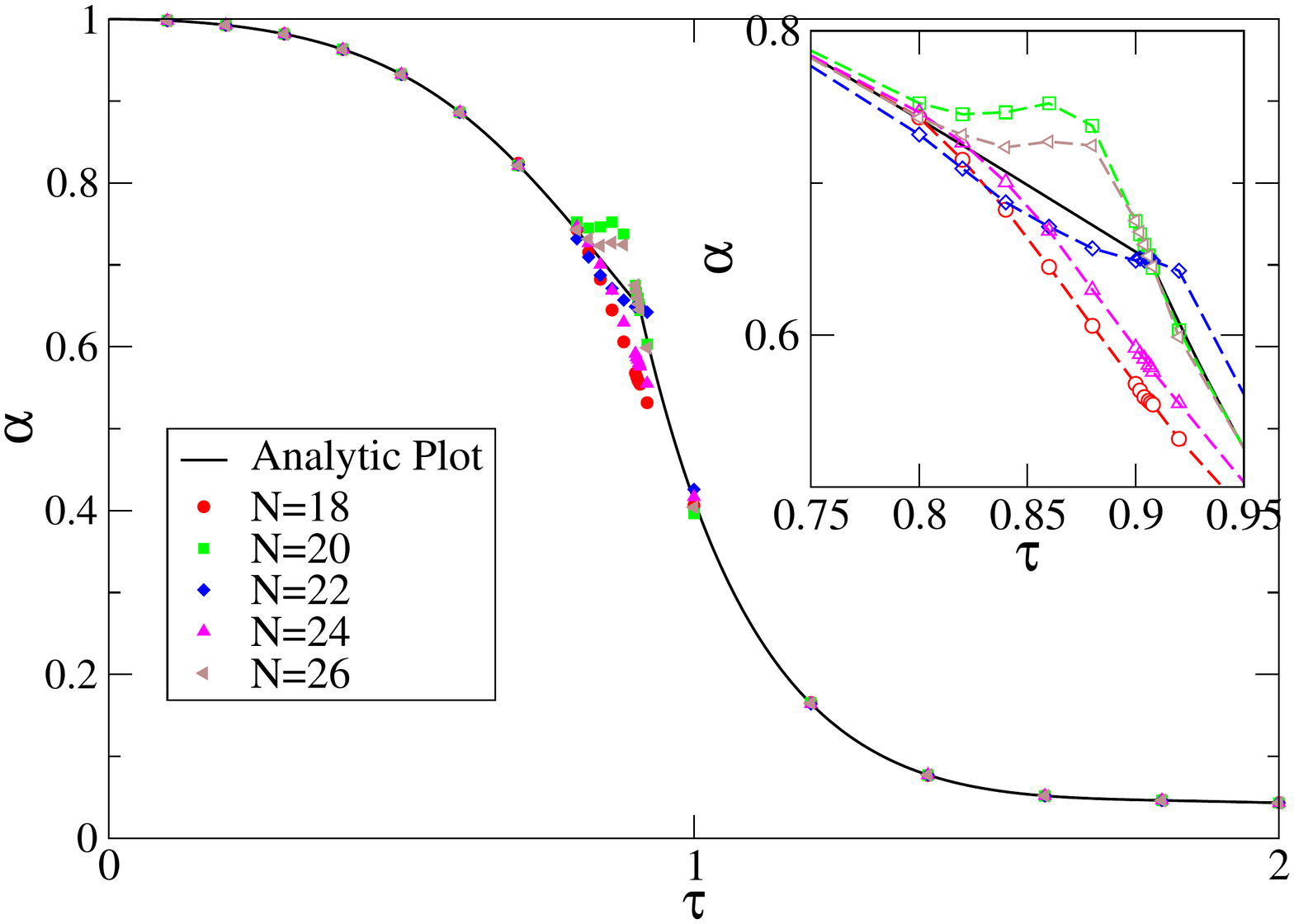}}
\caption{(color online) Decay constant $\alpha$ computed numerically for the measurement (\ref{Question1}), as a function of the applied transverse field $\Gamma$ (left panel) and as a function of $\tau$ (right panel) for finite system sizes. The left panel is for $\tau= 1.0$ and the right panel for $\Gamma =0.5$. The solid black curve is the analytic result from  Eq.~(\ref{alpha}). As in Fig. \ref{alpha_IQ=0}, the initial state corresponds to the ground state and we have checked that the results do not depend significantly on this choice.}
\label{smallN}
\end{figure*}

{\bf Other measurements - numerical study:}
When the measurement (\ref{Question1}) is replaced by a more general one
\be  {\emph Is} \;\; M_z = Q \,? \label{Question2} \ee
the states retained after each measurement is in general a superposition of several states $\sum_j C(j) |b_j\rangle$. 
It is difficult to proceed analytically now, since each state $|b_j\rangle$ after time evolution during the interval $\tau$ contribute to all other states resulting in a complicated interference phenomena. We have studied these cases numerically for small chains ($N \le 24$) and obtained the following results : (i)  For the measurement (\ref{Question2}), with $Q=0$, the decay constant $\alpha$ as a function of $\Gamma$  and $\tau$
does not  show any kink (Fig. \ref{alpha_IQ=0}) and is much smaller in magnitude than in the previous case (Fig. \ref{alpha_fig}).  Thus, the numerical results for small size preclude the existence of phase transition in this case.
 (ii) As mentioned above, for (\ref{Question1}) only the state with $M_z = 1$ is retained after each measurement and for (\ref{Question2}) all the states with $M_z \ne 0$ are retained. We have checked (numerically) that as we increase the number of states retained after each measurement, the peak in the curve for $\alpha$ gradually decreases. (iii) To check the reliability of our finite size numerical study, we have also applied it to the measurement (\ref{Question1}) (Fig.~\ref{smallN}). The decay constant does show a peak or kink at the critical values $\Gamma_0$ and $\tau_0$, but the position and height of the peak and kink vary strongly and non-monotonically with $N$.  Thus, the numerical study reproduces the analytic result presented in Fig. \ref{alpha_fig} and is reliable.

{\bf Effect of Initial State:} For the measurement of (\ref{Question1}), if we start at time $t=0$ with a wave function $|\psi(0)\rangle$, then the amplitude just after the first measurement will be $C_1 = \langle b_{N'} | \psi(0) \rangle $ according to Eqs. (\ref{collapse},\ref{amplitude}). However, the ratio of amplitudes as given by Eq. (\ref{ratioC}) is independent of $|\psi(0)\rangle$. Thus, although the individual probabilities, $p_n$ will depend on $n$, the ratio $p_n/p_0$ and hence the decay constant $\alpha$ is independent of the initial state. All the analytic results derived above for the  measurement of (\ref{Question1}), are therefore true for all initial conditions.\\
For the measurement of (\ref{Question2}) with $Q=0$, the numerical results show a small amount of variation with initial condition in an erratic way, but that issue is not consequential, since the absence of phase transition in this case is still maintained.

{\bf Discussions:}
We demonstrate analytically that periodic projective measurement of transverse magnetic moment on a transverse Ising chain shows a new type of phase transition. If we ask the question ``What is the probability that at the $n$-th measurement the total transverse magnetic moment $M_z$ will be not equal to $1$ for the first time?", then the decay constant of this probability is non-analytic as a function of transverse field strength and time-period of measurement. This non-analyticity is {\em not} a result of some singularity in ground state behaviour, but constitutes of contributions from all excited states and originates from the dynamics and the process of measurement itself. There is no physical process (like ordering or entanglement) or any symmetry breaking or scaling that accompanies this transition. After each measurement, a single state (in $k$ space) is retained, and during the time evolution thereafter, the wave function spreads over other states. The non-analyticity associated with this spreading is manifested as a phase transition. We also report that according to numerical investigations at small size, if we ask the question ``What is the probability that at the $n$-th measurement the total transverse magnetic moment $M_z$ will be zero for the first time?", then the decay constant of this probability does {\em not} show any singular behaviour.

We have investigated in this work the behaviour of transverse magnetisation, which is {\em not} the order parameter of the transverse Ising Hamiltonian. Although the phase transition we propose has no connection with the order-disorder phase transition, it would certainly be of interest to investigate the response of the system to a measurement related to the order parameter, namely, the longitudinal magnetisation. However, we could not include this study here, since the analytic calculation of longitudinal magnetisation is difficult to perform and meaningful results are obtained only from analytic studies. Work in this direction is in progress.

 Experimental study of cold atoms on optical lattices may provide an avenue for exploring such phenomena. A search for similar phase transitions for measurement of other quantities will also be of interest for theoreticians. 

{\bf Acknowledgement:} The authors are grateful to A. Dhar and D. Sen for helpful discussions. S. Dasgupta is grateful to ICTS-TIFR (Bangalore) for hospitality. One author (S. Dhar) is thankful to CSIR, India for the research fellowship (Grant no: 09/028(0839)/2011-EMR-I).


\begin{thebibliography}{99}
\bibitem  {formula} The sign function in this formula is often omitted in standard text books.
\end{thebibliography}
\end{document}